\documentclass[aps,pra,a4paper,nofootinbib,twocolumn,superscriptaddress]{revtex4}

\usepackage[english]{babel}

\usepackage[utf8]{inputenc}

\usepackage{amsmath,amssymb,amsfonts,mathrsfs}

\usepackage[amsmath,thmmarks]{ntheorem}

\usepackage{graphicx}

\usepackage{pdfpages}
\usepackage{wrapfig}
\usepackage{textpos}
\usepackage{subfigure}
\usepackage{float}

\usepackage{braket}
\usepackage{upgreek}
\usepackage{lipsum}

\usepackage{pgf}
\usepackage{tikz}

\usepackage[linkcolor=black,colorlinks=true,citecolor=black,filecolor=black]{hyperref}
\usepackage{dsfont}

\newcommand{\ii}{\mathrm{i}}

\newcommand{\E}[1]{\langle #1 \rangle}
\newcommand{\rbs}[1]{\left(#1\right)}
\newcommand{\sbs}[1]{\left[#1\right]}

\begin{document}

\title{Fluctuation dynamics of an open photon Bose-Einstein condensate}

\author{Fahri Emre Ozturk}
 \email[Email: ]{oeztuerk@iap.uni-bonn.de}
 \affiliation{Institut für Angewandte Physik, Universität Bonn, Wegelerstr. 8, 53115 Bonn, Germany} 

 \author{Tim Lappe}
\email[Email: ]{lappet@th.physik.uni-bonn.de}
 \affiliation{Physikalisches Institut and Bethe Center for Theoretical Physics, Universit\"at Bonn, Nussallee 12, 53115 Bonn, Germany}
 
\author{Göran Hellmann}
\affiliation{Institut für Angewandte Physik, Universität Bonn, Wegelerstr. 8, 53115 Bonn, Germany}

\author{Julian Schmitt}
\email[Present address: ]{Cavendish Laboratory, University of Cambridge, J. J. Thomson Avenue, Cambridge CB3 0HE, United Kingdom}
\affiliation{Institut für Angewandte Physik, Universität Bonn, Wegelerstr. 8, 53115 Bonn, Germany} 

\author{Jan Klaers}
\email[Present address: ]{Complex Photonic Systems (COPS), MESA+ Institute for Nanotechnology,\ \  University of Twente,\ \ 7522\ \  NB\ \ Enschede, The Netherlands}
\affiliation{Institut für Angewandte Physik, Universität Bonn, Wegelerstr. 8, 53115 Bonn, Germany}

\author{Frank Vewinger}
\affiliation{Institut für Angewandte Physik, Universität Bonn, Wegelerstr. 8, 53115 Bonn, Germany}    

 \author{Johann Kroha}
\affiliation{Physikalisches Institut and Bethe Center for Theoretical Physics, Universit\"at Bonn, Nussallee 12, 53115 Bonn, Germany} 

 
 \author{Martin Weitz}
 \affiliation{Institut für Angewandte Physik, Universität Bonn, Wegelerstr. 8, 53115 Bonn, Germany}    

\date{\today}

\begin{abstract}
Bosonic gases coupled to a particle reservoir have proven to support a regime of operation where Bose-Einstein condensation coexists with unusually large particle-number fluctuations. Experimentally, this situation has been realized with two-dimensional photon gases in a dye-filled optical microcavity. Here, we investigate theoretically and experimentally the open-system dynamics of a grand canonical Bose-Einstein condensate of photons. We identify a regime with temporal oscillations {\color{black}of the second-order coherence function $g^{(2)}(\tau)$}, even though the energy spectrum closely matches the predictions for an equilibrium Bose-Einstein distribution and the system is operated deeply in the regime of weak light-matter coupling. The observed temporal oscillations are attributed to the nonlinear, weakly driven-dissipative nature of the system which leads to time-reversal symmetry breaking.
\end{abstract}

\maketitle

\section{Introduction}\label{sec:intro}
Bose-Einstein condensates are the experimental basis of a variety of observed collective quantum phenomena \cite{Pitaevskii.2003,Bloch.2008}. Generally, Bose-Einstein condensation is a phenomenon in
thermodynamic equilibrium for Bose systems, usually described for a fixed total particle number,
which leads to a macroscopic ground-state occupation. In ultracold atomic gases \cite{Anderson.1995,Davis.1995}
and exciton-polariton systems \cite{Kasprzak.2006,Balili.2007}, condensation has been observed following thermalization by interparticle collisions, a process that leaves the total particle number constant. Photons usually do not exhibit condensation, with the chemical potential vanishing at all temperatures in the well-known example of a (three-dimensional) blackbody radiator. Two-dimensional photon gases under harmonic confinement can, however, reach condensation, as was experimentally demonstrated in dye-filled optical microcavities \cite{Klaers.2010b,Marelic.2015,Greveling.2018}. In this system, thermalization is reached by absorption and re-emission processes on the dye molecules, which leaves the average particle number constant but allows for fluctuations around this average value. {\color{black}In the limit of number fluctuations that become as large as the average particle number}, such a situation is described well by the grand canonical statistical ensemble \cite{Fujiwara.1970,Ziff.1977,Klaers.2012,Sobyanin.2012}. Lasers are long-known physical systems that also have macroscopic population of excited states, but which operate far from thermal equilibrium
\cite{Zamora.2017,Vorberg.2018}. The crossover from lasing to condensation,
characterized by a varying degree of thermalization,
has been investigated by observing deviations from a thermalized distribution,
both in polariton and photon gases
\cite{Deng.2010,Byrnes.2014,Sun.2017,Kirton.2013,Schmitt.2015,Marelic.2015}. Particle-number conserving condensates as well as lasers are characterized
by vanishing particle-number fluctuations in the thermodynamic limit,
i.e., by a value of the second-order coherence
function of $g^{(2)}(\tau)=1$ at all delay times $\tau$.

Experimentally, evidence for Bose-Einstein condensation in the grand canonical statistical regime with unusually large statistical number fluctuations has been observed with photons in the dye microcavity system \cite{Schmitt.2014}. The coupling of photons to the photo-excitable dye molecules implies that the dye does not only act as a heat bath, but also as a particle reservoir due to the possible interconversion of photons and dye electronic excitations. For a large relative size of the dye reservoir, this leads to strikingly enhanced statistical number fluctuations and a zero-delay second-order correlation $g^{(2)}(0) = 2$, i.e., as in a thermal source. Notably, these fluctuations, which can be as large as the average value, occur deep in the condensed phase. On the other hand, with a smaller effective relative size of the dye reservoir, the dye microcavity photon condensate can also be operated in the (usual) canonical statistical regime, with much smaller number fluctuations and a zero-delay intensity correlation $g^{(2)}(0)=1$. {\color{black}Independent of the ensemble conditions,} frequent collisions of solvent molecules with the dye on a timescale of $10^{-14}$ s \cite{Lakowicz.2006}, cause the dye microcavity condensate to operate in the weakly coupled regime of matter and light \cite{Angelis.2000,Yokoyama.1989}, i.e., the trapped particles are photons (not polaritons) and the system can be well described by a rate equation model, with, e.g., no Rabi oscillations occurring.

In the present work, we examine the temporal dynamics of a photon Bose-Einstein condensate (BEC) 
in the strongly fluctuating regime, where a steady driven-dissipative
state is induced by a balance of continuous dye pumping and cavity losses.
We observe distinct temporal oscillations of the photon number correlations
$g^{(2)}(\tau)$, even though the spectral photon distribution is, within experimental uncertainties, indistinguishable from predictions for thermodynamic equilibrium
(Bose-Einstein distribution). Temporal oscillations of density fluctuations
have been observed, e.g. in oscillatory relaxation dynamics in lasers
\cite{Takemura.2012} and {\color{black}in driven-dissipative atomic Bose-Einstein condensates confined in high-finesse optical cavities} \cite{Brennecke.2013}. Damped, oscillatory displacement
dynamics have been observed in colloids suspended in liquids upon
driving out of equilibrium \cite{Berner.2018}, see also theoretical work
proposing corresponding experiments with laser-driven quantum dots
\cite{Moradi.2018}. In all those systems, however, the (temporally averaged)
spectrum clearly differs from the thermodynamic equilibrium distribution.

For the correlation dynamics of the photon BEC, we find quantitative
agreement with a theoretical analysis in terms of nonlinear rate equations.
{\color{black} We are thus able to trace back the origin of the correlation oscillations
to the coupling of the dye reservoir and the condensate photons, combined with
time-reversal symmetry breaking in the (weakly) driven-dissipative state. The nonlinearity of the coupling implies that the dynamics of $g^{(2)}(\tau)$ will depend on the steady state of the system.}

The paper is organized as follows. Section \ref{sec:expsetup} provides
some details of the experimental setup and mode of operation.
In Section \ref{sec:theory} we derive the rate equations for the
nonequilibrium dynamics of the average numbers of photons and of
dye-molecule excitations as well as for
the autocorrelation functions of these quantities.
In Section \ref{sec:results} we present and discuss
the experimental results, along with the comparison to the theory. 
We conclude  in Section \ref{sec:conclusion} with an outlook to further studies.

\section{Experimental setup and mode of operation}\label{sec:expsetup}

\begin{figure}
	\centering
	\includegraphics[width=1.0\linewidth]{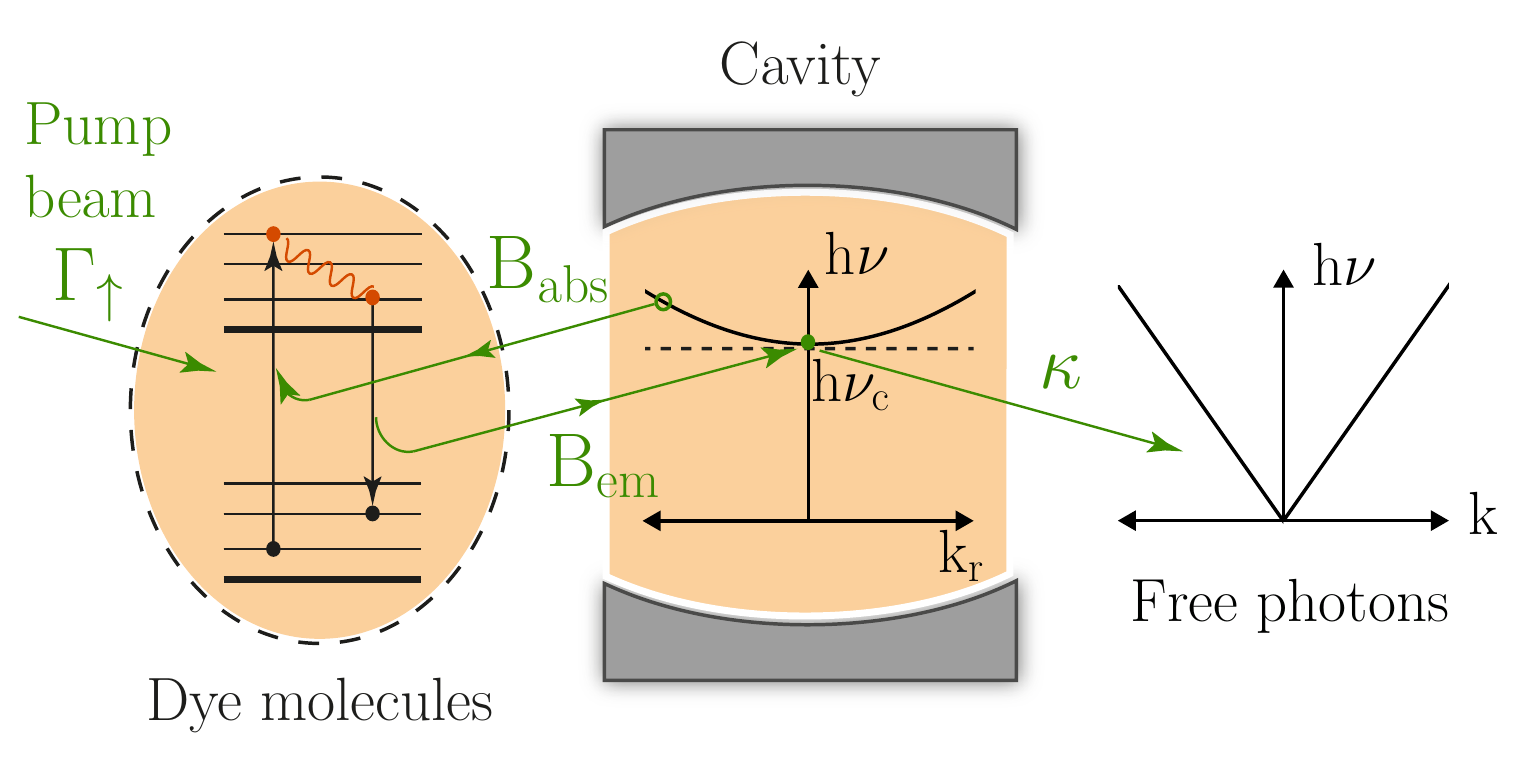}
	\caption{The photon flow through the condensed photon gas in the dye-filled microcavity. The dye molecules, which are
  excited with a pump rate $\Gamma_{\uparrow}$ constitute a heat and particle reservoir for the condensed photons. 
The photon dispersion in the cavity in terms of the transverse wave vector 
is $h\nu = \sqrt{(h\nu_c)^2+(\hbar c k_r)^2}$, with  $\nu_c =\omega_{c}/2\pi$ the 
cavity cutoff frequency and $c$ the light velocity in
the dye medium. The photon loss rate of the condensate through the cavity mirrors is represented by $\kappa$.}
\label{fig:1}
\end{figure}

Our experimental setup for trapping a two-dimensional photon gas in a dye-filled optical microcavity is similar to that described in earlier work \cite{Klaers.2010b,Klaers.2011}, as shown in Fig.~\ref{fig:1}. The cavity is composed of two highly reflecting mirrors (reflectivity $>99.998 \%$) of $1$ m curvature radius spaced by $\approx 2~\upmu$m distance, and is filled with rhodamine dye dissolved in ethylene glycol ($1$~mmol/L). Due to the small mirror spacing, the cavity has a longitudinal mode spacing comparable to the emission width of the dye. In this regime, we observe that only photons of a fixed longitudinal mode number $q$ populate the cavity, with $q=7$ in our setup. This imposes an upper limit on the optical wavelength and a restriction of energies to a minimum cutoff of $\hbar\omega_{c}\approx2.1$~eV for photons in the cavity, where $\omega_{c}$ is the cutoff frequency of the cavity with transverse momentum $k_r=0$. The optical dispersion becomes quadratic and the mirror curvature induces a trapping potential for the photons (see Fig.~\ref{fig:1}). One can show that the system is formally equivalent to a harmonically trapped two-dimensional gas of massive particles for which -- other than for (three-dimensional) blackbody radiation -- Bose-Einstein condensation is possible for a thermalized ensemble \cite{Klaers.2010b}.
To initially populate the cavity and compensate for losses from e.g.,
mirror transmission, the dye is pumped to a stationary state with an external laser beam.

To record the statistics as well as the number correlations of the photon condensate, the microcavity emission is directed through an optical telescope and a mode filter consisting of a $10.6$ mm diameter iris to separate the condensate mode from the higher-mode (thermal cloud) photons. {\color{black}Even though the divergence of the higher-order modes is larger than that of the ground mode, they contribute residually to the intensity near the optical axis in the far field, and a fraction of photons in these modes is consequently transmitted through the momentum filter. This imperfection of the mode filtering reduces the magnitude of the measured (normalized) intensity correlations. As a result, the measured correlations are smaller than the calculated values for the respective system parameters. We do not expect the temporal shape of the correlation signal to be affected by the imperfect mode filtering.}

The light passing through the momentum filter is then sent through a polarizer to remove the polarization degeneracy and then imaged onto a fast ($5$ GHz bandwidth) photomultiplier. The electronic output signal of the photomultiplier is analyzed employing a $3.5$ GHz bandwidth oscilloscope. To suppress the influence of electronic noise of the high-bandwidth electronic analysis system (which is mainly attributed to the oscilloscope’s analog-digital converters), the photomultiplier output is simultaneously recorded by two oscilloscope channels, with the cross-correlation used for further data analysis of the fluctuations. Calibration of the photomultiplier signal is performed via the measured spectra, which relate the photon number in the condensate peak to the known {\color{black}critical} photon number in the thermal photon cloud of   $N_{c}=\pi^2/3(k_{B}T/\hbar\Omega)^2=80660$ for the used experimental parameters ($T=300$ K, trap frequency $\Omega=2\pi\cdot40$ GHz).

\begin{figure}
 	\centering
 	\includegraphics[width=0.7\linewidth]{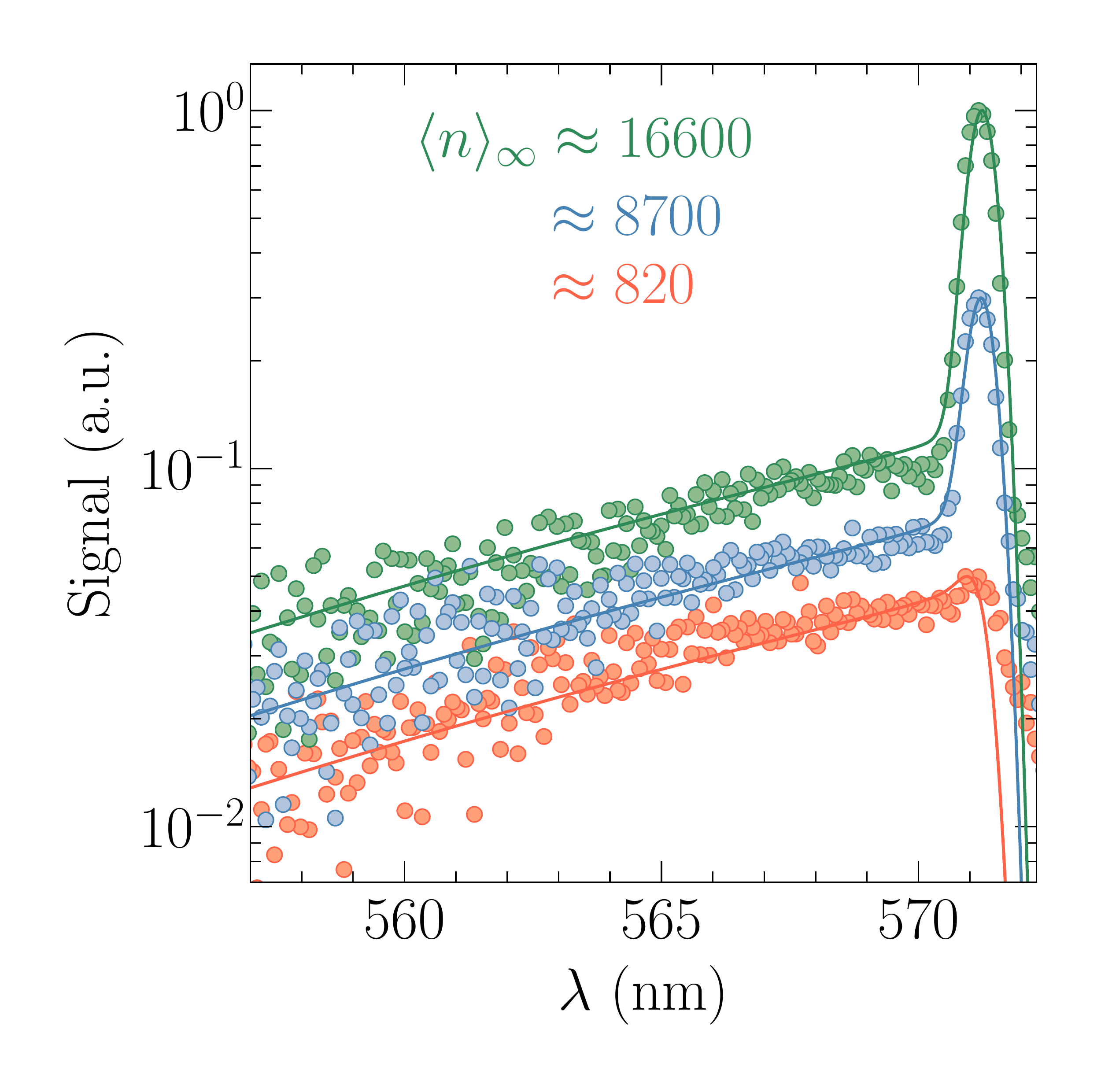}
 	\caption{Optical spectra (arbitrary units) as a function of 
          wavelength $\lambda$, taken from cavity emission through the mirrors
          at the cutoff wavelength of $\lambda_c = 571.3$ nm, where the
          experiment was conducted. The solid lines are fits of the
          Bose-Einstein distribution at $300$K (broadened by the experimental
          resolution) to the experimental data, with the steady-state average condensate photon number $\langle n\rangle_\infty$ as the only adjustable parameter. The fitted values of $\langle n\rangle_\infty$ are shown in the legend. The spectra at different $\langle n\rangle_\infty$ are vertically shifted for clarity.
          }
 	\label{fig:2}
 \end{figure}

The dephasing time $\tau_{\varphi}$ of dye excitations 
(given by the collision time of solvent molecules with the dye on a scale of $10^{-14}$ s is much shorter than the photon
lifetime in the cavity ($\tau_{\mathrm{p}}\approx 0.5$ ns,
determined by cavity losses), $\tau_{\varphi}\ll\tau_{\mathrm{p}}$. Furthermore, coherence between dye excitations and photons cannot be established, that is, our experiment operates in the weakly coupled regime of matter
and light, and the trapped particles are photons, not polaritons \cite{Angelis.2000,Yokoyama.1989}).
The photon lifetime, in turn, is much shorter than
the nonradiative decay time of dye excitations,
$\tau_{\mathrm{p}}\ll \tau_{\mathrm{nrad}}$ (with $\tau_{\mathrm{nrad}}$ being in the order of 50 ns \cite{Barroso.1998}.
This means that the sum of the number of cavity photons {\color{black}(including the thermal mode photons)} and the dye excitation number can be considered conserved on
the time scale of the photon lifetime $\tau_{\mathrm{p}}$, where the photon number alone is fixed on average only. {\color{black} To good approximation, also the sum of the number of condensate mode photons and dye excitations is conserved, as exchange between thermal cloud photons and condensate mode photons only occurs via the bath and the dye excitation number exceeds the (thermal mode) photon number by many orders of magnitude \cite{Klaers.2012}}
For sufficiently fast energy exchange with the dye reservoir, i.e., if
several absorption and re-emission cycles occur before a photon is lost,
the photons reach a thermal spectral distribution at the rovibrational
temperature of the dye, which is at room temperature ($300$~K). 
Fig.~\ref{fig:2} shows measurements of the optical spectrum for different average condensate photon numbers $\langle n\rangle_\infty$ (different pump rates). 
{\color{black}The visible good agreement of the the experimental data (dots) with the predictions of equilibrium theory (solid lines)} confirms that in our 
experiments, a thermal distribution is achieved to good approximation, 
despite continuous pumping and losses. 
Above a critical photon number, the thermal photon gas eventually 
forms a BEC, signaled by the condensate peak in the spectral distribution 
at the position of the low-frequency cavity cutoff on top of a broad 
thermal photon cloud \cite{Klaers.2010b}.

By varying the photon number $\langle n \rangle_\infty$
with respect to the number of dye reservoir molecules $M$ within the cavity volume, the photon number
statistics can be continuously tuned from a small relative reservoir size with Poissonian statistics and
$g^{(2)}(0)\approx 1$ to a strongly fluctuating state of small
relative photon number with Bose-Einstein statistics and $g^{(2)}(0)\approx2$ \cite{Schmitt.2014}.

\section{Theoretical description}\label{sec:theory}

\subsection{Model and master equation}

The thermalization dynamics of the
dye-filled microcavity has been studied  
\cite{Kirton.2013,Kirton.2015,Keeling.2016} employing a Frank-Condon model 
for the dye molecules: The two-level system of the electronic ground and 
excited states, between which the
optical transitions occur, is coupled to a molecular vibrational degree of
freedom (phonon) because the rest position of the ionic molecular oscillator
depends on whether the molecule is in its electronic ground or excited state. 
The total number of dye molecules is $M=M_\uparrow + M_\downarrow$, 
where $M_\uparrow$ and $M_\downarrow$ are the 
number of molecules in the electronically excited or ground state, 
respectively. The Hamiltonian for the dye-filled microcavity reads thus \cite{Marthaler.2011}
\begin{align}\label{eq:Ham}
    \begin{split}
        H &= \sum_{k}\omega_{k}^{\phantom{\dagger}} a^{\dagger}_{k}a^{\phantom{\dagger}}_{k} +  \sum_{m=1}^{M} \left[ \frac{\delta}{2}\sigma^z_m + \Omega\, b^{\dagger}_{m}b^{}_{m} \right. \\
        &+ \left.  \Omega\sqrt{S}\,\sigma^z_m\left( b^{}_{m} + b^{\dagger}_{m} \right)
         + g\sum_{k}\left(a^{\phantom{\dagger}}_{k}\sigma^+_m + a^{\dagger}_{k}\sigma^{-}_m \right) \right] ,
    \end{split}
\end{align}
in units such that $\hbar = 1$. The cavity-photon modes with transverse dispersion $\omega_k$
are represented by the bosonic operators $a^{}_{k}$, $a^{\dagger}_{k}$,
the vibronic states of dye molecule $m$ with oscillator
frequency $\Omega$ by $b^{}_{m}$, $b^{\dagger}_{m}$, and the electronic
two-level system of molecule $m$ by the Pauli matrix $\sigma^z_m$ and
and raising/lowering operators $\sigma^\pm_m$, with the electronic
transition frequency $\delta$. The Frank-Condon
electron-phonon coupling is parametrized by $S$,
where the phonon position operator is 
$\hat{x} \propto \left( b^{}_{m} + b^{\dagger}_{m} \right)$. 
The last term in Eq.~\eqref{eq:Ham} describes 
photon emission or absorption with the optical transition matrix element $g$,
the smallest energy scale in the system. 
Since we consider photon gases which have
already reached a stationary, thermal distribution, and since in the
experiment the transverse cavity ground-state mode (condensate mode) is
singled out, the analysis may be restricted to the photon correlations 
in the condensate, i.e., we will collapse the sum over cavity modes in
Eq.~\eqref{eq:Ham} to $k=0$, where $\omega_c \equiv \omega_0$ is the cavity-cutoff frequency (see also Fig. \ref{fig:1}). The cutoff is chosen such that the cavity detuning
is $\Delta = \omega_c - \delta < 0$. From here on, we will drop the cavity-mode subscript on the photon operators and write $a \equiv a_0$.

The molecular part of the Hamiltonian can be diagonalized by a polaron 
transformation \cite{Marthaler.2011}. This leads to an effective, 
nonlinear electron-photon coupling, mediated by the phonon excitations 
of the dye. Due to fast collisions of the dye molecules with solvent molecules, 
the phonon excitations may be considered to be in thermal equilibrium at 
ambient temperature.  After treating the
phonon excitations as a Markovian thermal bath, this coupling can be 
parametrized by a coherent part $g_\beta$ \cite{Radonjic.2018}, and incoherent,
phonon-assisted couplings $B_{\text{abs}}$ for photon absorption and $B_{\text{em}}$ 
for photon emission (see Fig.~\ref{fig:1}).
For the density matrix $\rho$, one obtains in this way the master equation 
\begin{align}\label{eq:master}
    \begin{split}
        \dot{\rho} = \ii\,[\rho, H_0 ] + \kappa\mathcal{L}[a^{}_{}]\rho 
         + \sum_{m=1}^M \mathcal{L}_m\rho,
    \end{split}
\end{align}
where $H_0 = \Delta a^\dagger_{}a^{}_{} + \sum_{m=1}^{M} g_\beta\rbs{a^\dagger_{}\sigma_m^- + a_{}^{}\sigma_m^+} $ is the Hamiltonian generating the coherent part of the evolution in the rotated frame. The 
parameter $\kappa$ in Eq.~\eqref{eq:master} describes the cavity loss to
the environment, where the Lindblad operator acting on the density matrix
$\rho$ is defined as $\mathcal{L}[X]\rho =
\tfrac{1}{2}\rbs{[X^{}_{}\rho, X^\dagger_{}] + [X^{}_{}, \rho X^\dagger_{}]}$.
The molecule-induced superoperator $\mathcal{L}_m$ is given by  
\begin{align}    \label{eq:Lmolecule}
    \mathcal{L}_m =  \Gamma_\uparrow\mathcal{L}[\sigma^+_m] +
    \Gamma_\downarrow\mathcal{L}[\sigma^-_m] +
    B_{\text{abs}}\mathcal{L}[a^{}_{}\sigma^+_m]  +
    B_{\text{em}}\mathcal{L}[a^\dagger_{}\sigma^-_m].
\end{align}
The four terms in $\mathcal{L}_m$ describe, 
in order of appearance, pumping by an external laser source 
($\Gamma_{\uparrow}$), nonradiative decay of dye excitations 
($\Gamma_{\downarrow}\ll\Gamma_{\uparrow}$), and photon absorption  
($B_{\text{abs}}$) and emission ($B_{\text{em}}$) by the dye molecules, respectively. 

Since the experiment operates in the regime $g_\beta/g \ll 1$ \cite{Radonjic.2018}, and the detuning is very large compared to the renormalized coherent coupling, $|\Delta| \gg g_\beta$, the cavity mode effectively couples to the molecules only incoherently via the Lindblad terms proportional to $B_{\text{abs}}$ or $B_{\text{em}}$. Therefore, the contribution of $H_0$ to the master
equation \eqref{eq:master} can be neglected for the present setup \cite{Kirton.2013}. In addition, one should note that 
$B_{\text{em}} > B_{\text{abs}}$ because of the red detuning of the cavity cutoff $\omega_c$ with respect to the electronic dye excitation energy $\delta$.

\subsection{Average particle numbers}

 The coupled rate equations for the average condensate mode photon number 
$\langle n \rangle$ and the number of dye molecules in excited states $\langle M_{\uparrow} \rangle$ can now be derived from the master equation, where $\langle \cdot \rangle = \mathrm{Tr}[\rho (t)\cdot]$ 
denotes the thermal and quantum mechanical average.
Inserting $\dot{\rho}$ from Eqs.~\eqref{eq:master} and \eqref{eq:Lmolecule}, using cyclic permutation under the trace, and 
$\sigma_m^\pm\sigma_m^\mp=(1\pm\sigma_m^z)/2$, leads to operator products
of $M_{\uparrow}=\sum_m(1+\sigma_m^z)/2$ and $n=a^{\dagger}_{}a^{}_{}$.
Again because of the fast dye-solvent collisions, coherent propagation 
of excitations of different dye molecules ($m'\neq m$) is negligible. 
This means that the sum over a large number of molecules amounts to 
an average, $\E{M_{\uparrow}}=\sum_m(1+\E{\sigma_m^z})/2=
M(1+\langle\sigma_m^z\rangle)/2$, and expectation values of 
higher-order operator products factorize, $\E{n M_\uparrow}\approx
\E{n}\E{M_\uparrow}$.
In this way, one obtains the nonlinear, coupled rate equations
\begin{subequations}\label{eq:mean-field}
\begin{align}
    \begin{split}\label{eq:n}
            \frac{\mathrm{d}}{\mathrm{d}t}{\langle n \rangle} &= -\kappa\E{n} - B_{\text{abs}}   \rbs{M - \E{M_\uparrow}} \E{n}\\
            &+B_{\text{em}}  \rbs{\E{n}  + 1} \E{M_\uparrow},
    \end{split}\\
    \begin{split}\label{eq:m}
        \frac{\mathrm{d}}{\mathrm{d}t}{\langle M_\uparrow \rangle} &= \Gamma_\uparrow\rbs{M - \E{M_\uparrow}} - \Gamma_\downarrow\E{M_\uparrow} \\
        &+ B_{\text{abs}}   \rbs{M - \E{M_\uparrow}} \E{n} - B_{\text{em}}  \rbs{\E{n}  + 1} \E{M_\uparrow}.
    \end{split}
\end{align}
\end{subequations}
These agree, in fact, with the semiclassical rate equations expected 
phenomenologically from pumping and nonradiative decay of molecule
excitations a well as stimulated and spontaneous photon emission into 
the cavity. 
Alternatively, one may solve the untruncated rate equations for $\E{n}$ and
$\E{M_{\uparrow}}$ together with three equations for the second moments
\cite{Kirton.2015}. This is discussed in detail in the Appendix.
For large $M$, both solution methods give the same results for $\E{n}$ and $\E{M_{\uparrow}}$ in the long-time limit. 
 
To calculate the steady-state second-order photon correlations 
in the next section, it will be necessary to have the average numbers of
photons and excited molecules in the steady state which is reached in the long-time limit, $\E{n}_\infty$ and $\E{M_\uparrow}_\infty$. 
This amounts to setting the time derivatives in Eqs.~\eqref{eq:mean-field} 
to zero, and one obtains for large molecule number $M\gg 1$, 
\begin{align}
    \E{n}_\infty &= \frac{M \left(B_{\text{em}} \Gamma_{\uparrow} - B_{\text{abs}} \Gamma_{\downarrow}\right)}{\kappa \left(B_{\text{em}} + B_{\text{abs}}\right)} + \mathcal{O}(1), \\
    \E{M_\uparrow}_\infty &= \frac{MB_{\text{abs}} + \kappa}{B_{\text{abs}} + B_{\text{em}}} + \mathcal{O}(1/M).
\end{align}
In our experiments, the pump rate strongly exceeds the nonradiative decay,
$\Gamma_\uparrow\gg\Gamma_{\downarrow}$, and $B_{\text{em}} + B_{\text{abs}} \approx B_{\text{em}}$. The ratio of emission and absorption is given by  $B_{\text{em}}/B_{\text{abs}} = \exp(-\hbar\Delta/k_BT)$, where $T$ is the phonon temperature (Kennard-Stepanov relation). With these simplifications, the steady-state photon number becomes approximately
\begin{align}
    \E{n}_\infty \approx \frac{M\Gamma_\uparrow}{\kappa}.
\end{align}
This expression is useful for converting $\E{n}_\infty$,
which is measured in the experiments, into the pump
parameter $\Gamma_\uparrow$ of the theoretical model and vice versa. When comparing to experimental data, however, a full numerical solution for the steady state of Eqs. \eqref{eq:mean-field} and \eqref{eq:variances} is used.

\subsection{Second-order correlation function}

The time-dependent photon density-density or second-order correlation function measured in the experiment is defined as 
\begin{align}
    \begin{split}
        g^{(2)}(\tau) &= \left.\frac{\E{n(t+\tau)n(t)}}{\E{n(t)}^2}\right|_{t\to\infty} = \frac{\mathrm{Tr}\left[a^\dagger_{}a^{}_{}\;\mathrm{e}^{\hat{L}\tau}\Tilde{\rho}_{\infty} \right]}{\mathrm{Tr}\left[ a^\dagger_{}a^{}_{}\rho_\infty  \right]^2 },
    \end{split}
\end{align}
where $\hat{L}$ is the total Liouvillian superoperator belonging to the master equation \eqref{eq:master}, $\rho_\infty = \lim_{t\to\infty}\rho(t)$ denotes the steady-state density matrix, and we define $\Tilde{\rho}_{\infty} := a^\dagger_{}a^{}_{}\rho_\infty $. Note in passing that, for the normal-ordered second-order correlation function, one would need to set $\Tilde{\rho}_{\infty} = a^{}_{}\rho_\infty a^\dagger_{}$.
Defining also an effective average $ \overline{X} := \mathrm{Tr} X \mathrm{e}^{\hat{L}\tau} \Tilde{\rho}_{\infty} $, one has $ g^{(2)}(\tau) = \overline{n}/ \langle n \rangle_\infty^2$. Formally, $\overline{n}$ and $\overline{M}_\uparrow$ obey almost the same definitions as $\langle n \rangle$ and $\langle M_\uparrow\rangle$, however with $\rho(t)$ replaced by
$\mathrm{e}^{\hat{L}\tau}\Tilde{\rho}_{\infty}$ and
$\mathrm{Tr}\,\Tilde{\rho}_{\infty} = \E{n}_\infty$. Thus, one finds
equations of motion analogous to Eqs. \eqref{eq:n} and \eqref{eq:m},
\begin{subequations}\label{eq:n_and_m_eff}
    \begin{align}
    \begin{split}
      \frac{\mathrm{d}}{\mathrm{d}\tau} \overline{n} &= -\kappa\overline{n} - B_{\text{abs}}  \overline{(M - M_\uparrow)n} \\
      &+ B_{\text{em}} \overline{(n+1)M_\uparrow},
    \end{split}\\
    \begin{split}
    \frac{\mathrm{d}}{\mathrm{d}\tau} \overline{M_\uparrow} &= \Gamma_\uparrow\left(\E{n}_\infty M - \overline{M_\uparrow} \right) - \Gamma_\downarrow\overline{M_\uparrow} \\
        &+ B_{\text{abs}}  \overline{(M - M_\uparrow)n} - B_{\text{em}} \overline{(n+1)M_\uparrow}.
    \end{split}
\end{align}
\end{subequations}
The {\color{black} cluster expansions} of averages of the higher-order operator products $\overline{n M_\uparrow}$ are truncated under the same conditions as discussed in the previous section (incoherent propagation of excitations of different molecules).
This amounts to letting $\overline{n M_\uparrow} = \E{n}_\infty \overline{M_\uparrow} + \E{n}_\infty \E{nM_\uparrow}_\infty + \E{M_\uparrow}_\infty \overline{n}- 2\E{n}_\infty^2 \E{M_\uparrow}_\infty$. By computing the steady-state density matrix numerically exactly for different molecule numbers of order $M \sim 10^2$, we have checked that {\color{black} the truncation is well justified already for intermediate system sizes, i.e. above threshold \cite{Kirton.2015}. The experiment is operated far inside this range of validity. Above threshold, we also have $\Gamma_\uparrow \gg \Gamma_\downarrow$, such that we may safely neglect the non-radiative molecule decay rate $\Gamma_\downarrow$ for the evaluations in Section \ref{sec:results}. }

In terms of the deviations of the second-order correlation functions from their relaxed values (attained for $\tau\to\infty$), 
\begin{align}
    \boldsymbol{g} = 
    \begin{pmatrix}
    \Delta g^{(2)}_n \\
    \Delta g^{(2)}_{n,\,M_\uparrow}
    \end{pmatrix} = 
    \begin{pmatrix}
    \overline{n}  - \langle n \rangle_\infty^2 \\
    \overline{M}_\uparrow - \langle n \rangle_\infty \langle M_\uparrow \rangle_\infty
    \end{pmatrix},
\end{align}
Eqs. \eqref{eq:n_and_m_eff} then become
\begin{subequations}\label{eq:g_2a}
    \begin{align}
    \begin{split}
    \frac{\mathrm{d}}{\mathrm{d}\tau}  \Delta g_n^{(2)} &= \E{n}_\infty\Big[-\kappa\E{n}_\infty - B_{\text{abs}}   \rbs{M - \E{M_\uparrow}_\infty} \E{n}_\infty\\
            &+B_{\text{em}}  \rbs{\E{n}_\infty  + 1} \E{M_\uparrow}_\infty\Big] - \kappa \Delta g_n^{(2)} \\
            &- B_{\text{abs}} \sbs{ (M - \E{M_\uparrow}_\infty)\Delta g_{n}^{(2)} - \E{n}_\infty\Delta g_{n,\,M_\uparrow}^{(2)} } \\
            &+ B_{\text{em}}\sbs{(\E{n}_\infty + 1)\Delta g_{n,\,M_\uparrow}^{(2)} + \E{M_\uparrow}_\infty \Delta g_{n}^{(2)}},
    \end{split}\\
    \begin{split}
      \frac{\mathrm{d}}{\mathrm{d}\tau}  \Delta g_{n,\,M_\uparrow}^{(2)} &=  \E{n}_\infty\Big[ \Gamma_\uparrow\rbs{M - \E{M_\uparrow}_\infty} - \Gamma_\downarrow\E{M_\uparrow}_\infty \\
        &+ B_{\text{abs}}   \rbs{M - \E{M_\uparrow}_\infty} \E{n}_\infty \\
        &- B_{\text{em}}  \rbs{\E{n}_\infty  + 1} \E{M_\uparrow}_\infty \Big] - (\Gamma_\uparrow + \Gamma_\downarrow) \Delta g_{n,\,M_\uparrow}^{(2)} \\
        &+ B_{\text{abs}} \sbs{ (M - \E{M_\uparrow}_\infty)\Delta g_{n}^{(2)} - \E{n}_\infty\Delta g_{n,\,M_\uparrow}^{(2)} } \\
        &- B_{\text{em}}\sbs{(\E{n}_\infty + 1)\Delta g_{n,\,M_\uparrow}^{(2)} + \E{M_\uparrow}_\infty \Delta g_{n}^{(2)}}.
    \end{split}
\end{align}
\end{subequations}
Using the steady-state solution of Eqs. \eqref{eq:mean-field}, one eventually finds a system of two coupled linear equations,
\begin{align}\label{eq:g_2}
    \partial_\tau\boldsymbol{g} = 
    \begin{pmatrix}
        -\kappa - \tilde{\Gamma}_{M}  & \tilde{\Gamma}_{n}  \\
        \tilde{\Gamma}_{M} & -\left(\Gamma_\uparrow +\Gamma_\downarrow\right) - \tilde{\Gamma}_{n}
    \end{pmatrix} \boldsymbol{g}.
\end{align}
The matrix elements are given by
\begin{align}\label{eq:matrix_elements}
    \begin{split}
        \tilde{\Gamma}_{M}  &= B_{\text{abs}} (M - \langle M_\uparrow \rangle_\infty) - B_{\text{em}}\langle M_\uparrow \rangle_\infty, \\
        \tilde{\Gamma}_{n} &= B_{\text{abs}} \langle n \rangle_\infty + B_{\text{em}}(\langle n \rangle_\infty + 1).
    \end{split}
\end{align}
The coupling constant $\tilde{\Gamma}_{M}$ is composed of an absorption term proportional to the number of ground-state molecules in the steady state ($M -  \E{M_\uparrow}_\infty$), and a corresponding emission term with the number of excited molecules. The coupling constant $\tilde{\Gamma}_{n}$ is given by an absorption term, and the terms corresponding to stimulated and spontaneous emission.

The result of Eq. \eqref{eq:g_2} is equivalent to what one would obtain from linearizing Eqs. \eqref{eq:mean-field} around the steady state and then applying the regression theorem.

\begin{figure}
 	\centering
 	\includegraphics[width=0.7\linewidth]{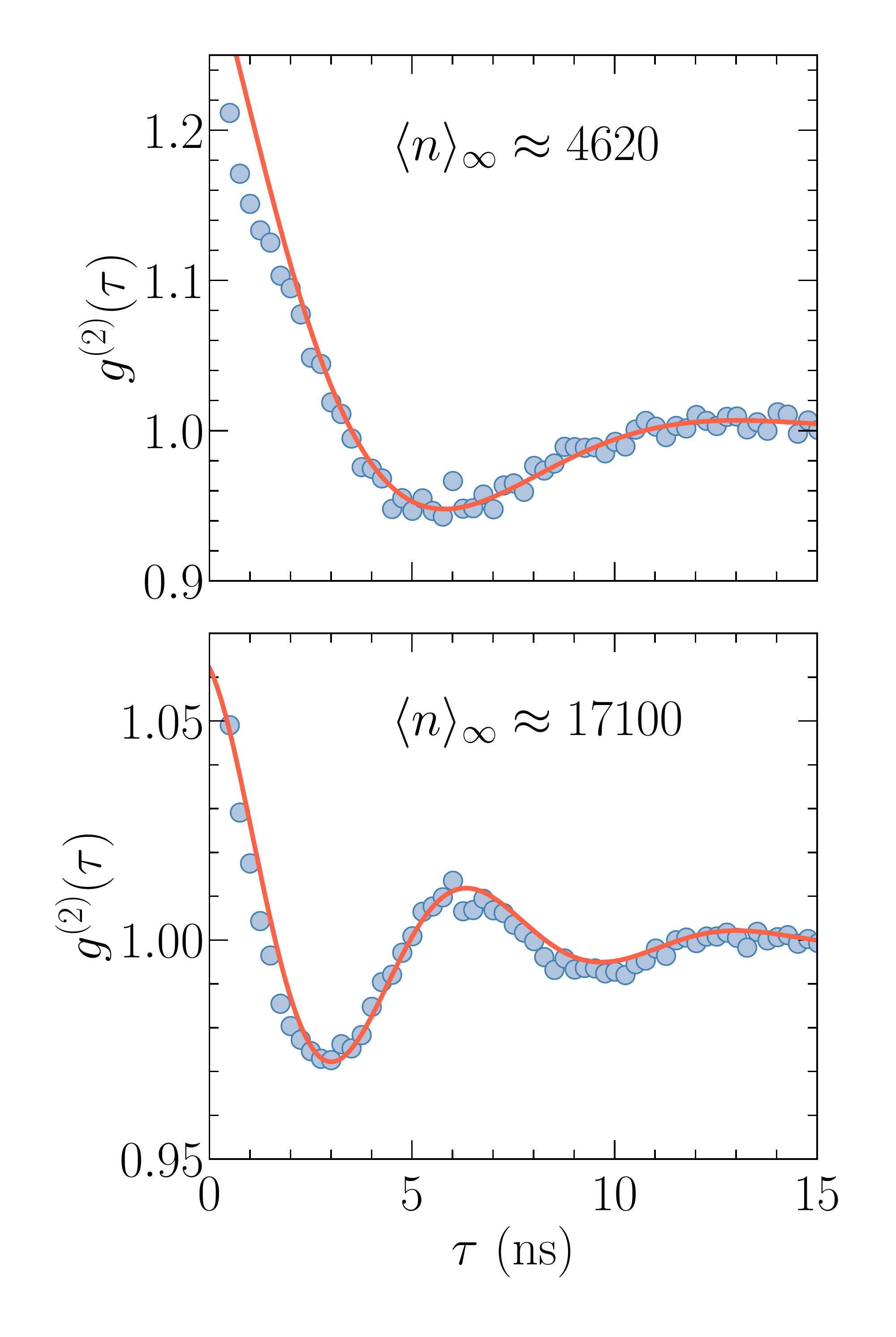}
 	\caption{Typical results for the time dependence of the second-order correlation function for two different photon numbers in the condensate, as indicated. {\color{black}The solid lines are fits to the model function, Eq. \eqref{eq:modelfunction}. The envelope of the observed correlation signal $g^{(2)}(\tau)$ temporally decays with a typical relaxation time of about 4 ns.}}
 	\label{fig:3}
 \end{figure}
 
 \section{Results}\label{sec:results}
 
The coupling matrix in Eq.~\eqref{eq:g_2} is non-Hermitian because of time-reversal
symmetry breaking in the driven-dissipative system. As a result, its
eigenvalues are found to be complex,
$\lambda = \lambda' \pm \ii\, \lambda''$, where $\lambda',\lambda'' \in\mathds{R}$. {\color{black} In terms of the quantities defined in Eqs. \eqref{eq:matrix_elements}, the eigenvalues are given by
\begin{align}
    \begin{split}
        \lambda &= -\gamma \pm \sqrt{\gamma^2 - \omega_0^2},
    \end{split}
\end{align}
where
\begin{subequations}
    \begin{align}
        2\gamma &= \kappa + \Gamma_\uparrow + \Gamma_\downarrow + \Tilde{\Gamma}_M + \Tilde{\Gamma}_n, \\
        \omega_0^2 &= \kappa \Tilde{\Gamma}_n + (\kappa + \Tilde{\Gamma}_M)(\Gamma_\uparrow + \Gamma_\downarrow).
    \end{align}
\end{subequations}
For $\Gamma_\downarrow = 0$, $B_{\text{abs}} = \alpha B_{\text{em}}$ and to leading order in $1/M$, this reduces to more insightful approximate expressions. In this case, we have
\begin{subequations}
    \begin{align}
        \E{n}_\infty &\approx M \frac{\Gamma_\uparrow}{(1+\alpha)\kappa}, \\
        \E{M_\uparrow}_\infty &\approx \frac{M\alpha B_{\text{em}} + \kappa}{(1+\alpha)B_{\text{em}}}.
    \end{align}
\end{subequations}    
This gives $\tilde{\Gamma}_{M}  = -\kappa$ and $\tilde{\Gamma}_{n} = M\Gamma_\uparrow B_{\text{em}}/\kappa$, from which we find
\begin{subequations}
    \begin{align}
        \gamma &= \frac{\Gamma_\uparrow}{2}\rbs{1 + \frac{MB_{\text{em}}}{\kappa}} \approx \frac{M\Gamma_\uparrow B_{\text{em}}}{2\kappa}, \\
        \omega_0^2 &= M\Gamma_\uparrow B_{\text{em}}.
    \end{align}
\end{subequations}    
Hence, the eigenvalues are given by
\begin{align}\label{eq:lambda}
    \begin{split}
        \lambda &= -\frac{M\Gamma_\uparrow B_{\text{em}}}{2\kappa} \pm \frac{1}{2}\sqrt{\rbs{\frac{M\Gamma_\uparrow B_{\text{em}}}{\kappa}}^2 - 4M\Gamma_\uparrow B_{\text{em}}}.
    \end{split}
\end{align}
Without drive and dissipation ($\Gamma_\uparrow = \kappa = 0$), we have a two-component system which shows a single relaxation time, i.e. the system possesses one zero eigenvalue and a purely real eigenvalue $\lambda = - ( \tilde{\Gamma}_{M} +  \tilde{\Gamma}_{n} )$. Correspondingly, the open-system character here is necessary to achieve an imaginary part for the eigenvalues and hence an oscillating photon-number correlation.}

{\color{black}Note that this approximation leading to Eq. \eqref{eq:lambda} is not employed in the analysis of the experimental data. It rather serves to illustrate how the openness of the system influences the eigenvalues through the steady-state occupations by making the dependence on $\Gamma_\uparrow$ and $\kappa$ explicit.}

\begin{figure}
    \centering
    \includegraphics[width=0.86\linewidth]{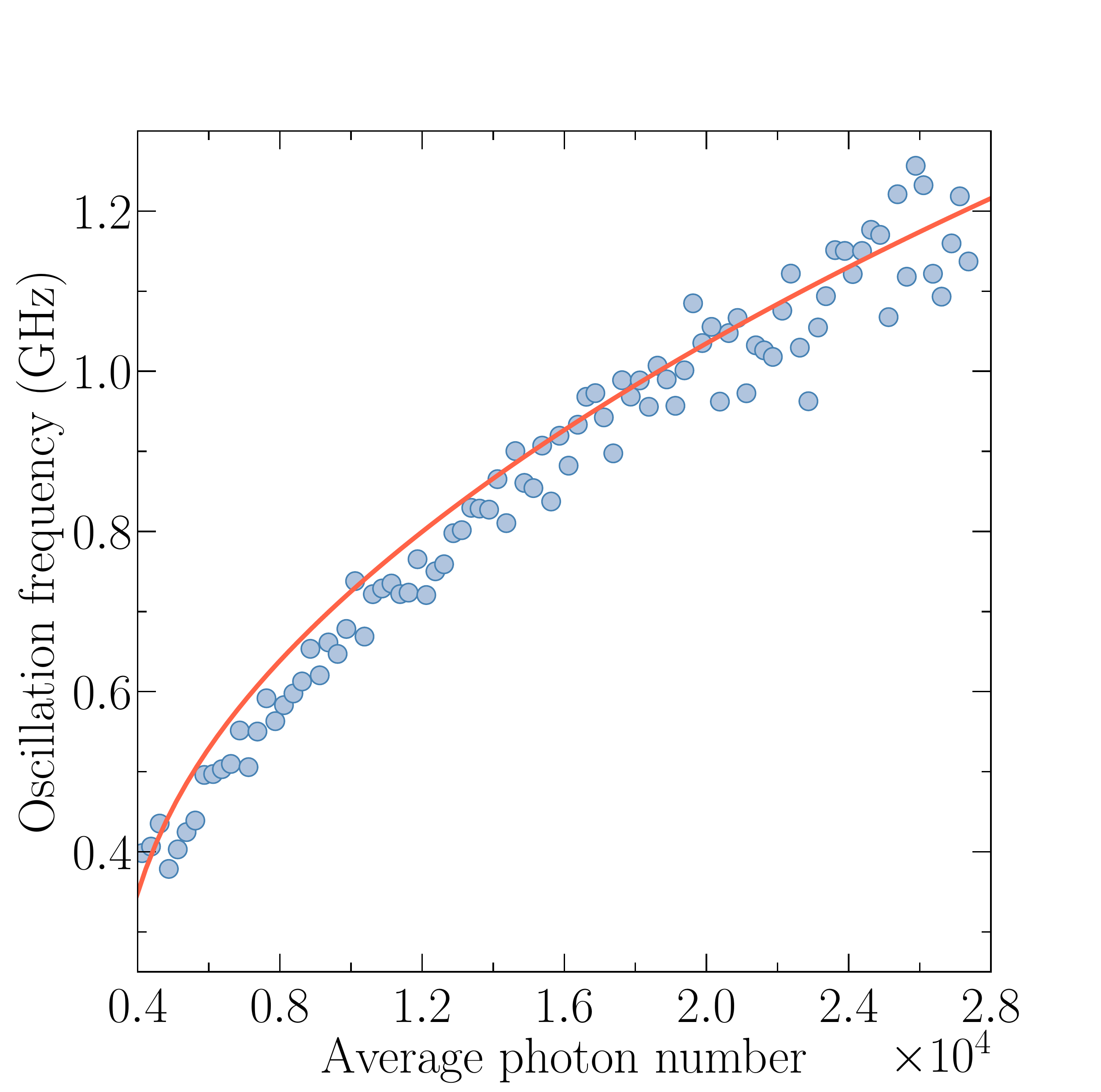}
    \caption{Oscillation frequency  $\omega^{(2)}$ of the second order correlation function $g^{(2)}(\tau)$ as a function of the average photon number  $\langle n\rangle_\infty$ in the condensate, as measured in the experiment (dots) and predicted by a nonlinear rate equation model (solid line). See the main text for details.  The parameter values are $M = 5.17\cdot 10^9$, $\kappa = 2.33$ GHz, $B_{\text{em}} = 2.50\cdot 10^{-5}$ GHz, $B_{\text{em}}/B_{\text{abs}} \approx 57$ (Kennard-Stepanov relation {\color{black}corresponding to the cutoff wavelength of $\lambda_c = 571.3$ nm}), $\Gamma_\downarrow = 0$.}
    \label{fig:4}
\end{figure}

{\color{black} For the second-order correlation function, one hence finds a solution of the form
\begin{align}\label{eq:modelfunction}
  g^{(2)}_n (\tau) = 1 + \mathrm{e}^{\lambda ' \tau} \sbs{ c_1 \cos({\lambda''\, \tau}) +  c_2 \sin({\lambda''\, \tau}) },
\end{align}
with the real part $\lambda ' <0$.} The initial values for the dynamics of the second-order correlation functions are found from the steady-state solutions for the second moments, $\E{n^2}_\infty$ and $\E{n M_\uparrow}_\infty$. Typical experimental data for the temporal variation of the second-order coherence function are given in Fig.~\ref{fig:3} for average photon numbers $\langle n \rangle_\infty\approx4620$ and $\langle n \rangle_\infty\approx17100$, respectively, showing damped, oscillatory behavior, as expected from the theoretical analysis. {\color{black}Within the studied experimental parameters, the oscillations persist as the system is tuned between grand canonical and canonical statistical regimes. As can be seen in the two examples of Fig.~\ref{fig:3}, the magnitude of photon number fluctuations (relative to the average photon number) decreases at higher average photon numbers corresponding to a smaller relative size of the dye reservoir. This behavior has been studied in more detail in earlier work \cite{Schmitt.2014}. For the present system, the theoretical values for $g^{(2)}(0)$ of the condensate mode at $\langle n \rangle_\infty\approx4620$ and $\langle n \rangle_\infty\approx17100$ are $\approx 2$ and $\approx 1.3$, respectively. As described above, the smaller $g^{(2)}(0)$ values observed in the experimental system are attributed to imperfect mode filtering, which tends to reduce the magnitude of the correlation signal towards that of an uncorrelated sample, where $g^{(2)}(\tau)=1$ for all times $\tau$. }{\color{black}The solid lines in Fig.~\ref{fig:3} are fits of the function from Eq. \eqref{eq:modelfunction} to the experimental data, with the imperfect mode filtering resulting in a lowering of the coefficients $c_1$,$c_2$ with respect to the theory values.} The experimental values of the second-order {\color{black} relaxation} time $\tau_c= 1/|\lambda '|$ and the oscillation frequency $\omega^{(2)}=\lambda ''$ of the $g^{(2)}$
correlations are determined by fitting the theoretical model function (Eq. \eqref{eq:modelfunction}) to the {\color{black} experimental data, as depicted} in Fig. \ref{fig:3}. In this way, we have recorded the variation of the oscillation frequency $\omega^{(2)}$ upon the change of the average photon number $\langle n\rangle_\infty$, as shown in Fig. \ref{fig:4} (dots). We observe an increase of the oscillation frequency of the second-order coherence function with the average photon number. The solid line in Fig. \ref{fig:4} is obtained using a fit of the theoretical eigenvalues
$\lambda '$, $\lambda''$ of Eq. \eqref{eq:g_2} to the experimental data,
where the model parameters $\kappa$, $\Gamma_{\uparrow}$,
and $B_{\text{em}}$ were used as fit parameters, and the nonradiative decay rate $\Gamma_{\downarrow}$ was set to zero.

The experimental data are fitted to good precision for all different $\E{n}_\infty$ by three parameters which are consistent with experimentally estimated values. We interpret this (as well as the comparison shown in Fig. \ref{fig:3}) as evidence that the
origin of the $g^{(2)}$ oscillations can be traced back to the effects
incorporated in our rate equation model, namely time-reversal symmetry
breaking due to nonequilibrium pumping and dissipation, and the coupling
between the subsystems of dye-molecule excitations and cavity photons.

{\color{black}Remarkably, despite the clear nonequilibrium
signatures in $g^{(2)}(\tau)$, a spectral distribution
which, within experimental accuracies, is indistinguishable from the
thermal equilibrium Bose-Einstein distribution is attained (see Fig.~\ref{fig:2}). The physical origin of this
seemingly contradictory behavior is that there is a (stationary) net flow
of photons from the dye reservoir to the resonator and out to the environment,
in such a way that the \textit{average} photon number is constant and the photon gas reaches a
thermal spectral distribution due to the thermal contact with the dye. Interestingly, our calculations show that the nonvanishing zero-delay second order correlations $g^{(2)}(0)>1$, from the grand canonical nature of the photon condensate, are  (in the investigated limits) not affected by the nonequilibrium character of the system. However, the residual photon loss and pump do affect the temporal decay of the correlation function $g^{(2)}(\tau)$. This photon flow causes additional structure in the intensity correlations, as observed in the present work. A related behavior is known in nanoelectronic systems, e.g. current carrying, metallic nanowires, where the electron energy distribution is equilibrated from  fast electron-electron collisions in the wire \cite{Steinbach.1996}. }

\section{Conclusions}\label{sec:conclusion}
To conclude, we observed an oscillatory behavior of grand canonical Bose-Einstein condensates by studying the second-order coherence of the emission of a dye microcavity. Its origin is traced back to the remnant driven-dissipative character of the light condensate. Our results show that even when the energy distribution of particles to good accuracy follows the predictions for thermal equilibrium, fluctuation dynamics depend sensitively on the openness of the system. {\color{black}We note that a related
behavior can be observed in the hot-electron regime of electronic quantum
wires at large bias voltage, where nonthermal noise, albeit
not oscillatory, is induced by the current. Due to fast electron-electron collisions in the wire, it coexists with an equilibrium (Fermi-Dirac) distribution of the electron energy. Here we observe this phenomenon
in a photon system for the first time.}

{\color{black} The $g^{(2)}(\tau)$ oscillations observed in our photon condensates are reminiscent of relaxation oscillations in lasers. However, there are important differences. First, a laser is in a state far from equilibrium with nonthermal spectral distribution. In contrast, our system is operated in a near-equilibrium state with a thermal Bose-Einstein distribution, but clearly nonequilibrium dynamics of the photon number correlations. Secondly, in a steady-state laser $g^{(2)}(\tau) = 1$ for all times. This means that oscillations of $g^{(2)}(\tau)$ do not occur unless the system is perturbed from outside, such as in the initial increase of the cavity photon number when a laser is turned on. For our system, when it is not operated with a very small relative size of the reservoir, the steady state is characterized by $g^{(2)}(0) > 1$. These fluctuations of the grand canonical system are responsible for the excitation of the oscillations.}

Our findings open up new avenues for further investigations of the open-system dynamics of grand canonical photon condensates. In the future, the experiments may be extended to further study the regime with stronger dissipation and drive. In addition, the second-order correlations can be used as a tool to sensitively characterize the system parameters. In a lattice with several coupled grand canonical photon condensates, a variety of new dynamical phases may be expected.

\textit{Note added in proof.} Recently, a  report on related experiments with
a dye-microcavity subject to short pulsed picosecond laser pump irradiation
appeared \cite{Walker2019}.

\section*{ACKNOWLEDGMENTS}
We acknowledge funding from the Deutsche Forschungsgemeinschaft (DFG) within the Cooperative Research Center SFB/TR 185 (277625399) and the Cluster of Excellence ML4Q (390534769), from the European Union within the ERC project INPEC and the Quantum Flagship project PhoQuS, and from the DLR within project BESQ.

\newpage

\onecolumngrid

\section*{APPENDIX: TRUNCATION OF HIERARCHY}\label{app:A}

In this Appendix, we present details on the hierarchy of the equations of motion
for the expectation values of successively increasing order. These will be calculated from the master equation \cite{Kirton.2013}
\begin{align}\label{eq:app_master}
    \begin{split}
        \dot{\rho} &=   \frac{\kappa}{2}\mathcal{L}[a^{}_{}]\rho + \frac{1}{2}\sum_{m=1}^{M}\Big\{\Gamma_\uparrow\mathcal{L}[\sigma^+_m] + \Gamma_\downarrow\mathcal{L}[\sigma^-_m] \\
        &+ B_{\text{em}}\mathcal{L}[a^\dagger_{}\sigma^-_m] + B_{\text{abs}}\mathcal{L}[a^{}_{}\sigma^+_m]\Big\}\rho.
    \end{split}
\end{align}
Assuming that the molecules are all identical, one can replace the sum over the $M$ molecules in the equation for the photon occupation $\E{a^\dagger_{}a^{}_{}}$ by a factor of $M$ and find
\begin{subequations}\label{eq:first-order}
    \begin{align}
        \begin{split}
                \partial_t{\langle a^\dagger_{}a^{}_{} \rangle} &= -\kappa\langle a^\dagger_{}a^{}_{} \rangle - \frac{M}{2}B_{\text{abs}} \langle a^\dagger_{}a^{}_{}  \left(1 -  \sigma^z \right)\rangle \\
                &+ \frac{M}{2}B_{\text{em}} \langle a^{}_{}a^\dagger_{}  \left(1 +  \sigma^z  \right) \rangle,
        \end{split}\\
        \begin{split}\label{eq:sigma_z}
            \partial_t{\langle \sigma^z \rangle} &= \Gamma_\uparrow\left(1 - \langle \sigma^z \rangle \right) - \Gamma_\downarrow\left(1 + \langle \sigma^z \rangle \right) \\
            &+ B_{\text{abs}} \langle a^\dagger_{}a^{}_{}  \left(1 -  \sigma^z \right)\rangle - B_{\text{em}} \langle a^{}_{}a^\dagger_{}  \left(1 +  \sigma^z  \right) \rangle,
        \end{split}
    \end{align}
\end{subequations}
where we have dropped the molecule index $m$. Multiplying Eq. \eqref{eq:sigma_z} by $M/2$ and using $M\partial_t \E{\sigma^z}/2 = \partial_t \E{M_\uparrow} $, $\E{\sigma^+\sigma^-} = \E{1 + \sigma^z}/2 = \E{m}/M $, $\E{\sigma^-\sigma^+} = \E{1 - \sigma^z}/2 = \E{M - m}/M $, and $\E{n M_\uparrow}\approx
\E{n}\E{M_\uparrow}$, we arrive at Eqs. \eqref{eq:mean-field}. The latter approximation is excellent for large systems: while     $\Delta g^{(2)}_{n,\,M_\uparrow}(0) = \langle M_\uparrow n\rangle_\infty - \langle n \rangle_\infty \langle M_\uparrow \rangle_\infty$ does not vanish in general, its influence on $\E{n}_\infty$, $\E{M_\uparrow}_\infty$ turns out to be negligible. 

Accordingly, the equations of motion of the next-order expectation values in Eqs. \eqref{eq:first-order} are given by
\begin{subequations}\label{eq:higher-order}
    \begin{align}
        \begin{split}
            \partial_t \langle a^\dagger_{} a^\dagger_{}a^{}_{}a^{}_{}\rangle &= -2\kappa \langle a^\dagger_{} a^\dagger_{}a^{}_{}a^{}_{}\rangle - 2 M B_{\text{abs}} \langle a^\dagger_{} a^\dagger_{}a^{}_{}a^{}_{} \sigma^-\sigma^+ \rangle
            + 2 M B_{\text{em}} \left(\langle a^\dagger_{} a^\dagger_{}a^{}_{}a^{}_{} \sigma^+\sigma^- \rangle + 2 \langle a^\dagger_{} a^{}_{} \sigma^+\sigma^- \rangle \right),
        \end{split}\\        
        \begin{split}\label{eq:a_dag-a-sig-sig}
                \partial_t{\langle a^\dagger_{}a^{}_{} \sigma^+\sigma^- \rangle} &= -\kappa\langle a^\dagger_{}a^{}_{}\sigma^+\sigma^-  \rangle + \Gamma_\uparrow \langle a^\dagger_{}a^{}_{} \sigma^-\sigma^+ \rangle  - \Gamma_\downarrow \langle a^\dagger_{}a^{}_{} \sigma^+\sigma^- \rangle  \\
        &+ B_{\text{abs}} \langle a^\dagger_{} a^\dagger_{}a^{}_{}a^{}_{} \sigma^-\sigma^+ \rangle 
        - B_{\text{em}} \left(\langle a^\dagger_{} a^\dagger_{}a^{}_{}a^{}_{} \sigma^+\sigma^- \rangle + 2\langle a^\dagger_{}a^{}_{}  \sigma^+\sigma^- \rangle\right)\\
        &- (M - 1)B_{\text{abs}}  \langle a^\dagger_{} a^{}_{} \sigma^+\sigma^- \Tilde{\sigma}^-\Tilde{\sigma}^+ \rangle
        + (M - 1)B_{\text{em}}  \langle a^{}_{}a^\dagger_{}  \sigma^+\sigma^- \Tilde{\sigma}^+\Tilde{\sigma}^- \rangle,
        \end{split}\\
        \begin{split}\label{eq:sig-sig-sig-sig}
            \partial_t \langle \sigma^+\sigma^- \Tilde{\sigma}^+\Tilde{\sigma}^- \rangle &= \Gamma_\uparrow \left( \langle \sigma^-\sigma^+ \Tilde{\sigma}^+\Tilde{\sigma}^- \rangle + \langle \sigma^+\sigma^- \Tilde{\sigma}^-\Tilde{\sigma}^+ \rangle\right) 
            - 2\Gamma_\downarrow \langle \sigma^+\sigma^- \Tilde{\sigma}^+\Tilde{\sigma}^- \rangle \\
            &+ B_{\text{abs}} \left( \langle a^\dagger_{}a^{}_{} \sigma^-\sigma^+ \Tilde{\sigma}^+\Tilde{\sigma}^- \rangle + \langle a^\dagger_{}a^{}_{} \sigma^+\sigma^- \Tilde{\sigma}^-\Tilde{\sigma}^+ \rangle\right) 
            - 2B_{\text{em}} \langle a^{}_{}a^\dagger_{} \sigma^+\sigma^- \Tilde{\sigma}^+\Tilde{\sigma}^- \rangle.
        \end{split}
    \end{align}
\end{subequations}
The Pauli matrices $\Tilde{\sigma}^\pm$ describe any molecule that is not identical to $\sigma^\pm$. Under the assumption that the total density matrix of the molecules is an incoherent mixture of all states corresponding to an excitation number of $M_\uparrow$ \cite{Kirton.2015}, which will be the case for the steady-state density matrix of the master equation \eqref{eq:app_master}, one can show that the expectation values of four Pauli matrices decompose as
\begin{align}\label{eq:trunc}
    \begin{split}
        M(M-1)\E{\sigma^+\sigma^-\Tilde{\sigma}^+\Tilde{\sigma}^-} &= \E{M_\uparrow^2} - \E{M_\uparrow}, \\
        M(M-1)\E{\sigma^+\sigma^-\Tilde{\sigma}^-\Tilde{\sigma}^+} &= M\E{M_\uparrow} - \E{M_\uparrow^2}.
    \end{split}
\end{align}
Then the truncation of Eqs. \eqref{eq:higher-order} can be performed rigorously by expanding the highest-order expectation values according to 
\begin{align*}
    \begin{split}
      & M(M-1)\E{a^{}_{}a^\dagger_{} \sigma^+\sigma^- \Tilde{\sigma}^+\Tilde{\sigma}^-} \\
      &= \E{(n+1)(M_\uparrow^2 - M_\uparrow)} \\
    &= \E{nM_\uparrow^2} - \E{nM_\uparrow} + \E{M_\uparrow^2} - \E{M_\uparrow},
    \end{split}
\end{align*}
and using the relation
\begin{align}
    \E{nM_\uparrow^2} = 2\E{M_\uparrow}\E{nM_\uparrow} + \E{n}\E{M_\uparrow^2} - 2\E{n}\E{M_\uparrow}^2.
\end{align}
In this manner, after multiplying Eq. \eqref{eq:a_dag-a-sig-sig} by a factor of $M$ and Eq. \eqref{eq:sig-sig-sig-sig} by $M(M - 1)$, one obtains \cite{Kirton.2015}
\begin{subequations}\label{eq:variances}
    \begin{align}\label{eq:n2}
    \begin{split}
        \partial_t\E{n^2} &= \partial_t\E{a^\dagger_{} a^{}_{}a^\dagger_{}a^{}_{}} =  \partial_t\E{a^\dagger_{} a^\dagger_{}a^{}_{}a^{}_{}} + \partial_t\E{a^\dagger_{}a^{}_{}} \\
        &= \kappa(\E{n} - 2\E{n^2}) - B_{\text{abs}}\left[ 2\E{n^2}(M - \E{M_\uparrow}) + 4\E{M_\uparrow}\E{n}^2 - 4\E{n}\E{nM_\uparrow} - \E{n(M - M_\uparrow)} \right] \\
        &+ B_{\text{em}}\left[ 4\E{n}\E{nM_\uparrow} + 2\E{M_\uparrow}\E{n^2} - 4\E{M_\uparrow}\E{n}^2 + \E{(3n + 1)M_\uparrow} \right],
    \end{split}\\
    \begin{split}
        \partial_t\E{nM_\uparrow} &= M\partial_t{\langle a^\dagger_{}a^{}_{} \sigma^+\sigma^- \rangle} \\
        &= -\kappa\E{nM_\uparrow} + \Gamma_\uparrow \E{n(M - M_\uparrow)}   - \Gamma_\downarrow \E{nM_\uparrow} \\
        &+ B_{\text{abs}}\left[ 2\E{M_\uparrow}\E{n}(\E{n} - \E{M_\uparrow}) - M(\E{n} + \E{nM_\uparrow}) + \E{n}\E{M_\uparrow^2} + 2(\E{M_\uparrow} - \E{n} + \tfrac{1}{2})\E{nM_\uparrow} + (M - \E{M_\uparrow})\E{n^2}\right] \\
        &+ B_{\text{em}} \left[ 2\E{M_\uparrow}\E{n}(\E{n} - \E{M_\uparrow}) - \E{M_\uparrow} + (\E{n} + 1)\E{M_\uparrow^2} + 2(\E{M_\uparrow} - \E{n} - 1)\E{nM_\uparrow} - \E{M_\uparrow}\E{n^2} \right],
    \end{split}\\    
    \begin{split}
            \partial_t \E{M_\uparrow^2} &= M(M - 1)\partial_t \langle \sigma^+\sigma^- \Tilde{\sigma}^+\Tilde{\sigma}^- \rangle + \partial_t\E{M_\uparrow}\\
            &= 2\Gamma_\uparrow \left( M\E{M_\uparrow} + M - \E{M_\uparrow} - \E{M_\uparrow^2}\right)
            +\Gamma_\downarrow \left(\E{M_\uparrow} - 2\E{M_\uparrow^2}\right) \\
            &+ B_{\text{abs}}\rbs{ (2M - 1)\E{nM_\uparrow} + M\E{n} - 4\E{M_\uparrow}\E{nM_\uparrow} - 2\E{n}\E{M_\uparrow^2} + 4\E{n}\E{M_\uparrow}^2 } \\
            &- B_{\text{em}}\rbs{ 4\E{M_\uparrow}\E{nM_\uparrow} + 2\E{n}\E{M_\uparrow^2} - 4\E{n}\E{M_\uparrow}^2 + 2\E{M_\uparrow^2} - (\E{n} + 1)\E{M_\uparrow} }.
    \end{split}
\end{align}
\end{subequations}
As mentioned in the main text, the steady-state solution of these equations is required for the initial values of the dynamics of the second-order correlation functions.

\vspace*{-0.45cm}

\twocolumngrid

\bibliographystyle{apsrev4-1}

%

\end{document}